\documentclass{aastex}
\usepackage{spr-astr-addons}
\usepackage{url}
\usepackage{psfig}   
\usepackage{subfigure}

\RequirePackage{color}

\begin{document}

\title{The effects of intense magnetic fields on Landau levels in a
neutron star }
\slugcomment{Not to appear in Nonlearned J., 45.}
\shorttitle{The electron Fermi energy }
\shortauthors{Z. F. Gao et al.}

\author{Z. F. Gao\altaffilmark{1,2,3}}
\altaffiltext{1}{Xinjiang Astronomical Observatory, CAS, 40-5 South Beijing Road,
Urumqi Xinjiang, 830011, China zhifu$_{-}$gao@uao.ac.cn}
\altaffiltext{2}{Graduate University of the Chinese
   Academy of Scienes, 19A Yuquan Road, Beijing, 100049, China}
\altaffiltext{3}{Department of Astronomy, Nanjing University, Nanjing, 210093, China}
\author{ N. Wang \altaffilmark{1}}
\affil{ Xinjiang Astronomical Observatory, CAS, 40-5 South Beijing Road, Urumqi
Xinjiang, 830011, China}
\author{ J. P. Yuan \altaffilmark{1}}
\affil{Xinjiang Astronomical Observatory, CAS, 40-5 South Beijing Road, Urumqi
Xinjiang, 830011, China}
\author{Chih-Kang. Chou \altaffilmark{4}}
 \affil{National Astronomical Observatories, Chinese Academy of Sciences, Beijing, 100012, China}

\begin{abstract}
In this paper, an approximate method of calculating the Fermi energy of
electrons ($E_{F}(e)$) in a high-intensity magnetic field, based on the analysis
of the distribution of a neutron star magnetic field, has been proposed.  In the
interior of a Neutron star, different forms of intense magnetic field could exist
simultaneously and a high electron Fermi energy could be generated by the release
of magnetic field energy.  The calculation results show that: $E_{F}(e)$ is related
to density $\rho$, the mean electron number per baryon $Y_{e}$ and magnetic field
strength $B$.
\end{abstract}


\section{Introduction}

The surface of a neutron star(NS) is widely thought to have magnetic
strengths as high as $10^{13}$ G.  The inner magnetic field is
assumed to be higher than that at the surface, and may be confined
within the crust or may be distributed throughout the entire
neutron star.  Thompson and Duncan (1996) predicted that
higher magnetic fields could exist in the interiors of these
so-called magnetars, which are strongly magnetized neutron stars
with field strengths several orders of magnitude greater than in
common radio pulsars.  Soft-gamma repeaters (SGRs) or anomalous X-ray
pulsars (AXPs) are considered as candidates for magnetars \citep{col00},
although this assumption requires further confirmation \citep{har99}.

For an ideal gas in equilibrium, the distribution function $f(E)$
can be expressed as
\begin{equation}
   f(E)=\frac{1}{exp[(E-\mu)/kT]\pm 1},
  \end{equation}
where the upper sign refers to fermions (Fermi-Dirac statistics),
the lower sign to bosons (Bose-Einstein statistics); $k$ and $\mu$
represent Boltzmann's constant and the particle chemical potential
(also called the Fermi energy $E_{F}$), respectively.  For a
completely degenerate fermion gas ($T\rightarrow 0$, i.e., $\mu/kT
\rightarrow \infty$), when $E\leq E_F, f(E) = 1$; when $E > E_F$,
$f(E) = 0$.  For a massive cooling neutron star, it is the pressure
associated with degenerate matter at zero temperature that supports
the entire star against gravitational collapse.  If all electrostatic
interactions are ignored, the electron gas could be treated as ideal
(noninteracting) at $T = 0$. $E_{F}(e)$ then has the simple form
\begin{equation}
E_{F}(e)=[p^{2}_{F}(e)c^{2}+ m^{2}_{e}c^{4}]^{\frac{1}{2}},
\end{equation}
with $p_{F}(e)$ the electron Fermi momentum.  In the interior of
a NS, when $B$ is too weak to be taken into consideration,
$p_{F}(e)$ is mainly determined by matter density (hereafter
referred to as $\rho$) and the mean electron number per baryon
(hereafter referred to as $Y_{e}$)\citep{sha83}.  The influence of
a high magnetic field on the equilibrium composition of a NS
has been shown in detail in previous studies \citep{yak01,lai91}.
On the basis of the distribution characteristics of Landau levels,
we introduce an approximate method for calculating the electron
Fermi energy, which holds in two different systems: an ideal
neutron-proton-electron (hereafter abbreviated to npe) gas and
the more realistic system of Baym, Pethick and Sutherland
(hereafter abbreviated to BPS) \citep{bay71} consisting of a
Coulomb lattice of heavy nuclei embedded in an electron gas
(in the crust), where the electron Fermi energy is very small
$\sim$(1-25) MeV corresponding to $\rho\sim (10^{4}- 10^{11})$
erg~cm$^{-3}$ \citep{bes93}.  Our method additionally holds in
two different magnetic fields: a weakly quantizing field and a
non-quantizing field.

This paper is organized as follows: in $\S$ 2, we summarize NS
structure and NS magnetic field, and consider the influences of
magnetic fields on NS matter and some simple non-equilibrium
processes; in $\S$ 3, we deduce an equation involving $E_{F}(e)$,
$\rho$, $B$ and $Y_{e}$, which is suitable for ultrastrong magnetic
fields; in $\S$ 4, a dispute on $E_{F}(e)$ in intense magnetic
fields is presented.  A brief conclusion is given in $\S$ 5.

\section{Effects of magnetic fields on neutron star matter}
This section is divided into three parts. To aid interpretation
of later sections, we review briefly the distribution of a NS
magnetic field and the relationship between the neutron star
magnetic field and Landau levels.  The details are as follows.

\subsection{Structure of a neutron star}
A NS can be subdivided into an atmosphere and four main internal
regions (the outer crust, the inner crust, the outer core, and
the inner core), where the electron Fermi energy grows with $\rho$.

The atmosphere is a thin layer of plasma which determines the
spectrum of thermal electromagnetic radiation of the star.
The geometrical depth of the atmosphere varies from some ten
centimeters in a hot star down to some millimeters in a cold one.
The outer crust, consisting of nuclei and electrons, extends from
the bottom of the atmosphere to the layer of density $\rho_{d}
\approx$ 4$\times10^{11}$ g~cm$^{-3}$, at which the Fermi energy
$E_{F}(n)= E_{F}(e)=$ 25 MeV, and has a depth of a few hundred
meters \citep{sha83}.  Owing to its relatively low density,
it contains about $10^{-5}$ of the total star mass. At
$\rho\leq10^{4}$ g~cm$^{-3}$, the electron gas may be non-
degenerate and the ionization may be incomplete; for $\rho <
10^{7}$ g~cm$^{-3}$, the ground state is $^{56}_{26}$Fe, the
nuclei capture electrons and become neutron-rich at density
$\sim 10^{9}$ g~cm$^{-1}$; and the neutrons start to drip from
the nuclei and form a free neutron gas at $\rho= \rho_{d}$.  The
inner crust, composed mainly of degenerate relativistic electrons
and non-relativistic nuclei over-saturated with neutrons, extends
from density $\rho_{d}$ at the upper boundary to $\sim$ 0.5$\rho_{0}$
at the base, and can be several kilometers deep, where $\rho_{0}$
= 2.8 $\times 10^{14}$ g~cm$^{-3}$ is the standard nuclear density.
When at the crust-core interface $\sim 1.4\times 10^{14}$ g~cm$^{-3}$,
the nuclei disappear completely.

The outer core, consisting of neutrons mixed with a small number
of protons and electrons, occupies the density range 0.5$\rho_{0}
\leq\rho\leq$ 2$\rho_{0}$ and has a depth of several kilometers.
When density approaches $\rho_{0}$, there are so many neutrons
that about ninety-five percent of particles are neutrons, with
only a small fraction of protons and electrons ($Y_{e}=Y_{p}\sim$
0.05)\citep{bec09, tsu02}. For $\rho \gg \rho_{0}$, where the
electron Fermi energy $E_{F}(e)> m_{\mu}c^{2} =$ 105.7 MeV, a
small fraction of muons ($\mu$) appear \citep{yak01}.  It is
worthwhile to note that a three-component liquid (neutrons,
protons and electrons) is in equilibrium under mutual $\beta-$
transformations.  The inner core is about several kilometers in
radius and has a central density as high as $\sim$ 10$\rho_{0}$.
For the inner core of a NS, with still further increase of density
it becomes energetically more economic if some nucleons transform
to `exotic' particles such as hyperons, pion condensates, kaon
condensates and quarks, etc, when $\rho> \rho_{tr},\rho_{tr}$ is
the transition density to these `exotic' particles, which is $\sim$
4$\rho_{0}$\citep{bec09, tsu02, tsu09}. The maximum of the inner
core density could exceed this transition density, so hyperons,
pion condensates, kaon condensates, quarks and nucleons with large
$Y_{e}$ are expected in the inner core of a NS \citep{tsu02, tsu09}.

\subsection{Magnetic fields of a neutron star}
Different forms of strong magnetic field could exist in the interior
of a magnetar at the same time.  For instance, in the absence of
superconductivity, the magnetic field is uniform on microscopic
scales, and many Landau levels are occupied by the particles
participating in neutrino reactions because the Fermi energies of
the particles are too high.  For example,  protons and electrons
occupy $\sim$ 300 Landau levels in a non-quantizing strong magnetic
field $\sim 10^{16}$ G at a density of about $1\sim $ several
$\rho_{0}$\citep{yak01}.  The neutrino emissivities in such cases
are about the same as in the non-magnetized matter because the
effects of magnetic quantization on the neutrino emissivities
are usually too weak.

On the contrary, if in the regions where protons are superconducting
and neutrons are superfluid, the magnetic field most likely exists
in the form of the quantized magnetic flux tubes (fluxoids), in such
case, the majority of the electrons are restricted to the ground Landau
level by this field referred to as strongly quantizing magnetic field;
in addition to these two types, an important phenomenon under study is
called weakly quantizing, which often occurs at low temperatures and high
densities, when only a few Landau levels are populated.

If the superhigh magnetic fields of magnetars originate from the
induced magnetic fields by the ferromagnetic moments of the
${}^3P_2$ Cooper pairs of the anisotropic neutron superfluid at a
moderate lower temperature ($T \ll 2.87\times10^{8}$ K, the
critical temperature of the ${}^3P_2$  neutron superfluid) and
high nuclear density ($\sim$ 0.5$\rho_{0} < \rho < $2.0 $
\rho_{0}$), then the maximum of magnetic field strength for the
heaviest magnetar may be estimated to be 3.0$\times 10^{15}$G
according to our model \citep{pen07,pen09}.

\subsection{Influences of magnetic fields on star matter}
The magnetic field greatly influences the properties of the
NS matter.  In particular, for the crust of a NS, when in a
strong magnetic field $B\sim 10^{12}$ G, individual atoms will
be elongated in the direction of this external field
\citep{flo77, flo79}.  In this case, crystals consisting of such
atoms differ substantially from normal crystals, so there must
clearly be anisotropy in their properties.  The influence of the
magnetic field on the crust matter structure of a NS is
determined by the parameter $\eta$\citep{rud71}.   The expression
for $\eta$ is
\begin{equation}
\eta = a_{0}/Za_{B}\simeq 15 B^{1/2}_{12}Z^{-3/2},
 \end{equation}
where $a_{0}$, $Z$, $a_{B}$ and $B_{12}$ are the first Bohr radius of
hydrogen atom, nuclear charge number, the radius of the electron cloud
in the quantizing magnetic field and the magnetic field in units of
$10^{12}$ G, respectively \citep{rud71}; $a_{B}$ can be expressed as
$a_{B}=(\hbar/m\omega_{B})^{1/2}$\citep{lan65}.  The external magnetic
field determines the shell structure of external electrons if $\eta\gg
Z^{-3/2}$.

A superhigh field ($B \geq B_{cr}$) can strongly quantize particle
motion, modify the phase space of protons (electrons), shift the beta-
equilibrium, increase the proton (electrons) fraction \citep{cha97,
lai91},  change the nuclear shell energies and nuclear magic numbers,
and therefore influence the nuclear composition and the equation of
states  of the inner crust of a NS.  Furthermore, magnetic fields
$\sim 10^{20}$ G may cause a substantial $n \rightarrow p$ conversion,
and as a result the system, composed of an ideal neutron-proton-
electron gas, may be converted to highly proton-rich matter
\citep{cha97}.  However, magnetic fields of such magnitude inside
NSs are uncertain and are also unauthentic.  To date, there have
been no observations indicating the existence of fields $B\geq
10^{16}$ G, in a NS interior; moreover, according to the virial
theorem, a magnetic field $B\geq 10^{18}$ G cannot exist in a NS
because the magnetic field energy ($\sim R^{3}B^{2}/6$) would
predominate over the gravitational binding energy ($\sim 3GM^{2}/5R$),
such that a dynamical instability in the hydrostatic configuration
would be induced by this ultrastrong magnetic field \citep{sha83,lai91}.

\section{Electron Fermi energy in superhigh magnetic fields}
We now consider a uniform magnetic field $B$ directed along the
$z$-axis.  In this case, in the Landau gauge the vector potential
$\overrightarrow{A}$ reads $\overrightarrow{A}= (-B_{y}, 0, 0)$.
For extremely strong magnetic fields, the cyclotron energy becomes
comparable to the electron rest-mass energy, and the transverse
motion of the electron becomes relativistic.  We can define a
relativistic magnetic field (often called a quantum critical
magnetic field $B_{cr}$) by the relation $\hbar\omega = m_{e} c^{2}$,
which gives $B_{cr} = m^{2} _{e}c^{3}/e\hbar$ = 4.414 $\times10^{13}$ G.
The electron energy levels may be obtained by solving the relativistic
Dirac equation in a strong magnetic field with the result
\begin{equation}
E_{e}=[m^{2}_{e}c^{4}(1+\nu\frac{2B}{B_{cr}})+p^{2}
_{z}c^{2}]^{\frac{1}{2}}
\end{equation}
where the quantum number $\nu$ is given by $\nu = n + \frac{1}{2}+
\sigma$ for the Landau level $n = 0, 1, 2, \cdots $, spin $\sigma
= \pm \frac{1}{2}$ \citep{can77}, and the quantity $p_{z}$ is the
$z$-component of the electron momentum and may be treated as a
continuous function. Combining $B_{cr} = m^{2}_{e}c^{3}/e\hbar $
with $\mu_{e}= e\hbar/2m_{e}c$ gives
\begin{equation}
  E_{e}^{2}= m_{e}^{2}c^{4}+p_{z}^{2}c^{2}
 +(2n+1+\sigma)2m_{e}c^{2}\mu_{e}B,
  \end{equation}
where $\mu_e\sim $0.927 $\times 10^{-20}$ ergs~G$^{-1}$ is the
magnetic moment of an electron.  According to the Pauli exclusion
principle, the electrons are situated in disparate energy states
in order one by one from the lowest energy state up to the Fermi
energy (the highest energy) with the highest momentum $p_{F}(z)$
along the magnetic field, and the electron energy state in a unit
volume, $N_{pha}$, should be equal to the electron number density,
$n_{e}$.  It is convenient to define a non-dimensional magnetic field:
$B^{*}= B/B_{cr}$ and the electron momentum perpendicular to
the magnetic field $p_{\perp} = m_{e}c\sqrt{(2n+1+\sigma)B^{*}}$.
Using the relation $2\mu_{e}B_{cr}/m_{e}c^{2} = 1$ and summing
over electron energy states in a 6-dimension phase space, we can
express $N_{pha}$ as follows:
\begin{eqnarray}
 &&N_{pha}=\frac{2\pi}{h^{3}}\int dp_{z}\sum_{n=0}^{n_{m}
 (p_z,\sigma,B^{*})}\sum g_{n}\nonumber\\
 &&\int \delta(\frac{p_{\perp}}{m_{e}c}-[(2n+1+\sigma)B^{*} ]
 ^{\frac{1}{2}}) p_{\perp}dp_{\perp},
\end{eqnarray}
where $\delta(\frac{p_{\perp}}{m_{e}c}-[(2n+1+\sigma)B^{*}]^{\frac{1}
{2}})$ is the Dirac $\delta$-function.  For $n = 0$, the spin is
antiparallel to $B$, the spin quantum number $\sigma = -1$, so the
ground state Landau level is non-degenerate; whereas at higher levels
$n > 0$ are doubly degenerate, and the spin quantum number $\sigma =
\pm 1$.  Therefore the spin degeneracy $g_{n} = 1$ for $n = 0$ and $g_{n}
 = 2$ for $n \geq 1$, then Eq.(5) can be rewritten
\begin{eqnarray}
 &&N_{pha}=2\pi(\frac{m_{e}c}{h})^{3}\int d(\frac{p_{z}}{m_{e}c})
 [\sum_{n=0}^{n_{m}(p_z,\sigma,B^{*})}\nonumber\\
 &&\int \delta(\frac{p_{\perp}}{m_{e}c}-(2nB^{*} )^{\frac{1}{2}})
 (\frac{p_{\perp}}{m_{e}c})d(\frac{p_{\perp}}{m_{e}c})
 +\sum_{n=1}^{n_{m}(p_z,\sigma,B^{*})}\nonumber\\
 &&\int \delta(\frac{p_{\perp}}{m_{e}c}-(2(n+1)B^{*})^{\frac{1}{2}})
 (\frac{p_{\perp}}{m_{e}c})d(\frac{p_{\perp}}{m_{e}c})],
\end{eqnarray}
where the maximum $z$-momentum $p_{F}(z)$ is defined by
\begin{equation}
 [p_{F}(z)c]^{2} + m^{2}_{e}c^{4}+(2n+1+\sigma)m^{2}_{e}
 c^{4}B^{*}\equiv E^{2}_{F}(e).
\end{equation}
The maximum Landau level number $n_{m}$ is the upper limit of the
summation over $n$ in Eq.(7), which is uniquely determined by the
condition $[p_{F}(z)c]^{2}\geq 0$ \citep{lai91}.  The expression for
$n_{m}$ is

\begin{equation}
 n_{m}(\sigma=-1)=
 \newline
 Int[\frac{1}{2B^{*}}[(\frac{E_{F}(e)}{m_{e}c^{2}})
 ^{2} -1 -(\frac{p_{z}}{m_{e}c})^{2}]],
\end{equation}

\begin{equation}
 n_{m}(\sigma=1)=
 \newline
 Int[\frac{1}{2B^{*}}[(\frac{E_{F}(e)}{m_{e}c^{2}})
 ^{2} -1 -(\frac{p_{z}}{m_{e}c})^{2}]- 1],
 \end{equation}
where $Int[x]$ denotes an integer value of the argument
$x$. Eq.(7) may now be rewritten
\begin{eqnarray}
 &&N_{pha}=2\pi(\frac{m_{e}c}{h})^{3}\int_{0}^{\frac{E_{F}(e)}{m_{e}c^{2}}}
[\sum_{n=0}^{n_{m}(p_z,\sigma=-1,B^{*})}\nonumber\\
 &&\sqrt{n}+\sum_{n=1}^{n_{m}(p_z,\sigma=1,B^{*})}\sqrt{n+1}]\sqrt{2B^{*}} d(\frac{p_{z}}{m_{e}c}).
    \end{eqnarray}
The term $\sum_{n=1}^{n_{m}(p_z,\sigma=1, B^{*})}\sqrt{n+1}$ can be
treated as
\begin{eqnarray}
&&\sum_{n=1}^{n_{m}(p_z,\sigma=1,B^{*})}\sqrt{n+1}
=\sum_{n=0}^{n_{m}(p_z,\sigma=1,B^{*})}\sqrt{n+1}-\nonumber\\
&&(\sqrt{1}+\sqrt{0})=\sum_{n'=0}^{n'_{m}(p_z,\sigma=1,B^{*})}\sqrt{n'} - 1,
\end{eqnarray}
where
\begin{eqnarray}
&&n'_{m}(p_z,B^{*},\sigma=1)= n_{m}(p_z,B^{*},\sigma=-1)\nonumber\\
&&=Int[\frac{1}{2B^{*}}[(\frac{E_{F}(e)}{m_{e}c^{2}})^{2}-1 -
(\frac{p_{z}}{m_{e}c})^{2}]].
\end{eqnarray}
Then we have
\begin{eqnarray}
&&N_{pha}=2\pi(\frac{m_{e}c}{h})^{3}\int_{0}^{\frac{E_{F}(e)}{m_{e}c^{2}}}
\sqrt{2B^{*}}[2\sum_{n=0}^{n_{m}}\sqrt{n} \nonumber\\
&&-1]d(\frac{p_{z}}{m_{e}c})=4\pi(\frac{m_{e}c}{h})^{3}\int_{0}^{\frac{E_{F}(e)}
{m_{e}c^{2}}}\sqrt{2B^{*}}[2\sum_{n=0}^{n_{m}}\nonumber\\
&&\sqrt{n} -\frac{1}{2}]d(\frac{p_{z}}{m_{e}c})
=4\pi(\frac{m_{e}c}{h})^{3}\sqrt{2B^{*}}\int_{0}^{\frac{E_{F}(e)}{m_{e}c^{2}}}\nonumber\\
&&\frac{2}{3}n^{\frac{3}{2}}_{m}d(\frac{p_{z}}{m_{e}c})
-2\pi(\frac{m_{e}c}{h})^{3}\sqrt{2B^{*}}(\frac{E_{F}(e)}{m_{e}c^{2}}).
\end{eqnarray}

Note, in the crust or interior of a NS, if the density
is so high that the electron longitudinal kinetic energy
exceeds its rest-mass energy, or if the magnetic field is
so high that the electron cyclotron energy also exceeds its
rest-mass energy, the electron becomes relativistic in either
case.  We introduce a ratio $q$ defined as $q = I_{1}/I_{2}$,
where $I_{1}=\int_{0}^{n_{m}}\sqrt{n}dn$ and $I_{2}=\sum_{n=0}
^{n_{m}}\sqrt{n}$.  If we assume $n_{m}$ to be 5, 6, 7, 8, 9, 10,
15, 20 and 30, then the corresponding values of $q$ are 0.889,
0.905, 0.916, 0.925, 0.932, 0.938, 0.957, 0.967 and 0.977, respectively.
It is easy to see that $q$ increases with $n_{m}$ and  $q\simeq 1$
if $n\gg 1$.  Therefore, when $n_{m}(p_z, B^{*}) \geq$ 6, the summation
formula can be approximately replaced by the following integral
equation:
\begin{equation}
\sum_{n=0}^{n_{m}}\sqrt{n}\simeq
\int_{0}^{n_{m}}\sqrt{n}dn =
\frac{2}{3}n^{\frac{3}{2}}_{m}.
\end{equation}
For simplicity, we focus on the crustal regions where the matter
density is high, the magnetic field $ B^{*}\leq 1$ (see $\S $
2) and Eq.(15) holds approximately.  Substituting Eq.(14) into Eq.(13)
we obtain
\begin{eqnarray}
&&N_{pha}=6\pi\sqrt{2B^{*}}(\frac{m_{e}c}{h})^{3}\int_{0}^{\frac{E_{F}(e)}
{m_{e}c^{2}}}n^{\frac{3}{2}}_{m}\nonumber\\
 &&d(\frac{p_{z}}{m_{e}c})-2\pi(\frac{m_{e}c}{h})^{3}
 \sqrt{2B{*}}(\frac{E_{F}(e)}{m_{e}c^{2}})\nonumber\\
&&=6\pi\sqrt{2B^{*}}(\frac{m_{e}c}{h})^{3}(\frac{1}{2B^{*}})^{\frac{3}{2}}
\int_{0}^{\frac{E_{F}(e)}{m_{e}c^{2}}}[(\frac{E_{F}(e)}{m_{e}c^{2}})^{2}-1\nonumber\\
&&-(\frac{p_{z}}{m_{e}c})^{2}]^{\frac{3}{2}}d(\frac{p_{z}}{m_{e}c})
 -2\pi(\frac{E_{F}(e)}{m_{e}c^{2}})(\frac{m_{e}c}{h})^{3}\sqrt{2B^{*}}\nonumber\\
 &&=\frac{3\pi}{B^{*}}(\frac{m_{e}c}{h})^{3}\int_{0}^{\frac{E_{F}(e)}
{m_{e}c^{2}}}[(\frac{E_{F}(e)}{m_{e}c^{2}})^{2} -1
 -(\frac{p_{z}}{m_{e}c})^{2}]^{\frac{3}{2}}\nonumber\\
 &&d(\frac{p_{z}}{m_{e}c})-2\pi(\frac{E_{F}(e)}{m_{e}c^{2}})
 (\frac{m_{e}c}{h})^{3}\sqrt{2B^{*}}.
\end{eqnarray}

In order to derive the formula for $E_{F}(e)$, we firstly introduce
two non-dimensional variables $\chi$ and $\gamma_{e}$, which are
defined as $\chi=(\frac{p_{z}}{m_{e}c})/(\frac{E_{F}(e)} {m_{e}c^{2}})
= p_{z}c/E_{F}(e)$ and $\gamma_{e}=E_{F}(e)/m_{e}c^{2}$, respectively,
then Eq.(16) can be rewritten as
\begin{eqnarray}
&& N_{pha}=\frac{3\pi}{B^{*}}(\frac{m_{e}c}{h})^{3}(\gamma_{e})^{4}
 \int_{0}^{1}(1-\frac{1}{\gamma^{2}_{e}}\nonumber\\
 &&-\chi ^{2})^{\frac{3}{2}}d\chi-2\pi\gamma_{e}
 (\frac{m_{e}c}{h})^{3}\sqrt{2B^{*}}.
\end{eqnarray}
The electron number density is determined by
\begin{equation}
      n_{e}  =  N_{A}\rho Y_{e},
   \end{equation}
where $N_{A}=$ 6.02$\times 10^{23}$ is the Avogadro constant \citep{sha83}.
For a given nucleus with proton number $Z$ and nucleon number $A$, the
relation $Y_{e}= Z/A$ always holds.  Combining Eq.(17) with Eq.(18), we have
\begin{eqnarray}
&&\frac{3\pi}{B^{*}}(\frac{m_{e}c}{h})^{3}(\gamma_{e})^{4}\int_{0}^{1}
(1-\frac{1}{\gamma^{2}_{e}}-\chi ^{2})^{\frac{3}{2}}d\chi\nonumber\\
&&-2\pi\gamma_{e}(\frac{m_{e}c}{h})^{3}\sqrt{2B^{*}}=N_{A}\rho Y_{e},
\end{eqnarray}
where $1/\gamma^{2}_{e}$ is called the modification factor.  The
calculation shows that the value of $1/\gamma^{2}_{e}$ decreases
with increasing $E_{F}(e)$, and the value of $1/\gamma^{2}_{e}$
is too small to be included in the calculation when $E_{F}(e) \geq
$ 5 MeV. Eq.(19) can be rewritten as
\begin{equation}
\frac{(3\pi)^{2}}{16B^{*}}(\gamma_{e})^{4}-2\pi\gamma_{e}\sqrt{2B^{*}}
=(\frac{m_{e}c}{h})^{-3}N_{A}\rho Y_{e},
\end{equation}
where the relation $\int_{0}^{1}(1-\chi^{2})^{\frac{3}{2}}d\chi=
3\pi/16$ is used.  Now, we simply discuss the application conditions
of Eq.(15).  It is important to note that, in a
non-relativistic weak field, the electron cyclotron energy is $\hbar
\omega_{B}= \hbar eB/(m_{e}c)$= 11.5 $B_{12}$ KeV, the maximum Landau
level number $n_{m}\sim E_{F}(e)/\hbar\omega_{B}\sim 10^{2}$ or higher,
where $B_{12}$ is magnetic field in units of $10^{12}$ G; also, in the
case of a weakly quantizing relativistic strong magnetic field ($B\sim
10^{14}\sim 10^{15}$ G), the solution of non-relativistic electron
cyclotron motion equation $\hbar\omega_{B}$ is no longer suitable,
but if this equation is used, the rest mass of an electron $m_{e}$
must be replaced by its effective mass $m^{*}_{e}$, which is far
larger than the former after taking into account the effect of
relativity.  In this latter case $n_{m}$ could be estimated to be
$\sim 10$ or higher, rather than 0 or 1, which shows our
evaluations are reasonable.  In fact, all of the analytic
derivations in this Section are based on the solution of the
relativistic Dirac equation for the electrons.  So the range of
Eq.(15) could be $B^{*}\sim$(0.1-100) and $\rho \sim$ $(10^{6}\sim 10^{15})$ g~cm$^{-3}$.

For a NS, the magnetic field varies with density, and the
properties of star matter are greatly influenced by the magnetic
field.  Details of how magnetic fields influence the structure and
the properties of a NS have been discussed briefly in $\S$2.3. However,
we can understand such influences gradually through experiments, due
to the peculiar circumstances of NSs (such as ultrastrong magnetic
fields and superhigh densities etc). If we take into account the
influence of a strong magnetic field on a cold, neutral gas of free
electrons and a single species of nucleus ($Z, A$), then applying
the BPS system based on a semi-empirical nuclear mass formula can
yield the relationship for $\rho$ and $B$.  The equilibrium nuclei
and the maximum equilibrium densities for specific
nuclei are listed in Table 1 for $B^{*}= 0, 0.1, 1, 10, 100$.

\begin{table*}[t]
\small
\caption{Magnetic BPS equilibrium nuclei below neutron drip. \label{tb1}}
\begin{tabular}{@{}crrrrr@{}}
\tableline
\multicolumn{1}{c}{$\rm Name$\tablenotemark{b}}&\multicolumn{1}{c}{$\rho_{max1}$\tablenotemark{a}} & \multicolumn{1}{c}{$\rho_{max2}$\tablenotemark{a}} &
 \multicolumn{1}{c}{$\rho_{max3}$\tablenotemark{a}}& \multicolumn{1}{c}{$\rho_{max4}$\tablenotemark{a}} &\multicolumn{1}{c}{$\rho_{max5}$\tablenotemark{a}}  \\
          & ($B^{*}= 0$)         & ($B^{*}$ = 0.1) &  ($B^{*}$ = 1)  &  ($B^{*}$= 10) &  ($B^{*}$= 100)    \\
\tableline
$^{56}_{26}$Fe~~...............................& $ 7.99 \times 10^{6}$ & $ 8.01 \times
10^{6} $ & $ 9.06 \times 10^{6}$ & $ 4.84 \times 10^{7}$ & $4.67 \times 10^{8}$  \\

$^{62}_{28}$Ni~~...............................& $ 2.71 \times 10^{8}$ &
& $ 2.72 \times 10^{8}$ & $ 3.10 \times 10^{8}$ & $ 1.68 \times 10^{9}$  \\

$^{64}_{28}$Ni~~...............................& $ 1.32 \times 10^{9}$ &
&                    & $ 1.34 \times 10^{9}$ & $ 2.78 \times 10^{9}$\\

$^{66}_{28}$Ni~~...............................& $ 1.54 \times 10^{9}$ & &    & $ 1.52 \times 10^{9}$ &NO \tablenotemark{c}\\

$^{86}_{36}$Kr~~...............................& $ 3.11 \times 10^{9}$ & &    &      & $ 3.87 \times 10^{9}$\\

$^{84}_{34}$Se~~...............................& $ 9.98 \times 10^{9}$ &
&                    &                    & $ 1.14 \times 10^{10}$\\

$^{82}_{32}$Ge~~...............................& $ 2.08 \times 10^{10}$ &
&                    &                    & $ 2.10 \times 10^{10}$\\

$^{80}_{30}$Zn~~...............................& $ 5.91 \times 10^{10}$ &
&                    &                    & $ 5.69 \times 10^{10}$\\

$^{78}_{28}$Ni~~...............................& $ 8.21 \times 10^{10}$ &
&                    &                    & $ 8.13 \times 10^{10}$\\

$^{126}_{44}$Ru~~............................... & $ 1.19 \times 10^{11}$ &
&                    &                    & $1.20 \times 10^{11}$\\

$^{124}_{42}$Mo~~............................... & $ 1.66 \times 10^{11}$ &
&                    &                    & $ 1.67 \times 10^{11}$\\

$^{122}_{40}$Zr~~...............................& $ 2.49 \times 10^{11}$ &
&                    &                    & $ 2.50 \times 10^{11}$\\

$^{120}_{38}$Sr~~............................... & $ 3.67 \times 10^{11}$ & & & & \\

$^{122}_{38}$Sr~~...............................& $ 3.89 \times 10^{11}$ & & & & \\

$^{118}_{36}$Kr~~...............................& $ 4.41 \times 10^{11}$ & & & & \\

\tableline
\end{tabular}
\tablenotetext{a}{The `a' symbol indicates that $\rho_{max}$ is the maximum density at which
the nuclide is present.}
\tablenotetext{b}{The `b' symbol indicates that the first six nuclear masses are known
experimentally.  The remainder are from \\
the Janecke-Gravey-Kelson
mass formula (see \citep{wap76}).}
\tablenotetext{c}{ The `c' symbol indicates that $^{66}_{28}$Ni is found to be absent from the
equilibrium nucleus sequence, so is not presented.}
\tablecomments{This Table is cited from \citep{lai91}.  All the values of $\rho_{max}$ are
measured in g~cm$^{-3}$.}
\end{table*}

In Table 1, when $B\neq 0$, the nucleus transition densities
become indistinguishable in the leading three digits to those that
are field-free, so the transition densities no longer continue to
be listed.   Table 1 clearly shows that a high-intensity magnetic
field alters the nucleus transition densities for the low-$A$
nuclei.  For the highest field $B$ = 100 $B_{cr}$, $^{62}_{28}Ni$
is found to be absent from the equilibrium nucleus sequence.  As
the density increases, the nuclei become increasingly saturated
with neutrons, but in all cases, neutron drip occurs at  $\rho
=\rho_{d}$.  Now, for the purpose of reducing error, an
approximate method for calculating $E_{F}(e)$ ($E_{F}(e)\leq$ 5
MeV) is introduced.  For instance, if we want to obtain the value
of $E_{F}(e)$ corresponding to the equilibrium nucleus density
$\rho$ = 1.34$\times 10^9$ g~cm $^{-3}$ for the nucleus $^{64}_{28}$Ni
when the magnetic field $B^{*}$= 10, we should firstly calculate
the value of the integral, which is written as follows:
\begin{eqnarray}
&&\int_{0}^{1}(1-\frac{1}{\gamma^{2}_{e}}-\chi^{2})^{\frac{3}{2}}d\chi \nonumber\\
&& \simeq\int_{0}^{1}(1-(\frac{0.511}{4.31})^{2}-\chi
^{2})^{\frac{3}{2}}d\chi=0.5726,
\end{eqnarray}
where the value of $E_{F}(e)\sim$ 4.31 MeV corresponding to
$B^{*}$= 0 is used.  Simplifying Eq.(20) gives
\begin{equation}
\frac{3\pi}{1}\times
0.5726(\gamma_{e})^{4}-2\pi\gamma_{e}\sqrt{2}
=(\frac{m_{e}c}{h})^{-3}N_{A}\rho Y_{e}.
\end{equation}
Inserting the values $Y_{e}$ = 0.4375, $\rho$ = 1.34$\times
10^9$g~cm$^{-3}$, $m_{e}$ = 9.11$\times10^{-28}$ g, $h$ =
6.63$\times10^{-27}$ erg ~s, $c= 3\times10^{10}$ cm s$^{-1}$
and $N_{A}$ = 6.02$\times10^{23}$ into Eq.(22) yields $\gamma
_{e}\sim$ 9.964, then $E_{F}(e)$ is estimated to be 0.511
$\times$ 9.964 MeV $\approx$ 5.09 MeV.  It's worthwhile to
note that the value of $E_{F}(e)$ is $\sim$ 5.05 MeV if
modification factor $1/\gamma^{2}_{e}$ is ignored.  From
Eq.(20), it's clear that $E_{F}(e)$ is a function of
$B$, $Y_{e}$ and $\rho$. Combining Table 1 with Table 2 can
allow us to calculate $E_{F}(e)$ in any given intense magnetic
field, the results are partly shown as follows.

 \begin{table*}[t]
\small
\caption{ The relation of $E_{F}(e)$ and $B$.\label{tb2}}
\begin{tabular}{@{}crrrrr@{}}
\tableline
 Name   & $Y_{e}$  &$E_{F}(1)$  & $E_{F}(2)$ &$E_{F}(3)$& $E_{F}(4)$    \\
          &          & $B^{*}$=0 &  $B^{*}$=1 &  $B^{*}$=10 &  $B^{*}$=100    \\
$ ^{56}_{26}$Fe & 0.4643  & 0.95\tablenotemark{a}&1.07\tablenotemark{b} &   2.53\tablenotemark{c}  &  6.93\tablenotemark{d}     \\
\tableline
$ ^{62}_{28}$Ni & 0.4516    & 2.61\tablenotemark{a}&   1.96\tablenotemark{b}& 3.62\tablenotemark{c}& 9.48\tablenotemark{d}       \\

$ ^{64}_{28}$Ni  & 0.4375    & 4.31\tablenotemark{a}&             & 5.09\tablenotemark{c} & 10.66\tablenotemark{d}     \\

$ ^{66}_{28}$Ni  & 0.4242    & 4.45\tablenotemark{a} &             &5.17\tablenotemark{c}  &  NO\tablenotemark{e}\\

$ ^{86}_{36}$Kr & 0.4186    & 5.66\tablenotemark{a}&             &             & 11.45\tablenotemark{d} \\

$ ^{84}_{34}$Se  & 0.4048    & 8.49\tablenotemark{a}&              &       & 14.88\tablenotemark{d}   \\

$ ^{82}_{32}$Ge   & 0.3902    &11.44\tablenotemark{a}&             &        & 17.18\tablenotemark{d}     \\

$ ^{80}_{30}$Zn    & 0.3750    & 14.08\tablenotemark{a}&            &        & 21.82\tablenotemark{d}   \\

\tableline
\end{tabular}
\tablenotetext{a}{
The sign `a' denotes that these values of $E_{F}(e)$ are \\
known experimentally, corresponding to nuclei densities  \\
7.96 $\times10^{6}$, 2.71 $\times10^{8}$, 1.30 $\times10 ^{9}$,
1.48 $\times10^{9}$, 1.48 $\times10^{9}$, \\
1.48 $\times10^{9}$, 1.10 $\times10^{10}$, 2.08$\times10^{10}$ and
5.44$\times10^{10}$g~cm$^{-3}$ \\
, respectively. Each density is the maximum density at  \\
which a given nucleus survives (Hanensel \& Pichon 1994);}
\tablenotetext{b}{
The sign `b' denotes that these nuclear masses known \\
experimentally are from the fourth column of Table 1;}
\tablenotetext{c}{
The sign `c' denotes that these nuclear masses known \\
experimentally are from the the fifth column of Table 1;}
\tablenotetext{d}{
The sign `d' denotes that these nuclear masses known \\
experimentally are from the sixth column of Table 1;}
\tablenotetext{e}{
The sign `e' denotes that $^{66}_{28}$Ni is found to be absent\\
from the equilibrium nucleus sequence, and so is not\\
presented.}
\tablecomments{ All the values of $E_{F}(e)$ are measured in MeV.  \\
Despite the existence of very small mass differences \\
caused by experimental uncertainty between the data  \\
in the second column of Table 1 and that in the third \\
column of Table 2, the results of our calculations will\\
not be affected.   In the fourth column of Table 2, for \\
every nucleus ( from $^{64}_{28}$Ni to $^{80}_{30}$Zn ), the discrepancy\\
between the nucleus transition density corresponding \\
to $B^{*}=1$ and the nucleus transition density corres-\\
ponding to $B^{*} = 0$  is so small that it is almost imp- \\
ossible to make a significant difference.   The same is  \\
true, in the fifth column of Table 2, for every nucle- \\
us (from $^{86}_{36}$Kr to $^{80}_{30}$Zn), the discrepancy between the\\
nucleus transition density corresponding to $B^{*}= 10$ \\
and the nucleus transition density corresponding to \\
$B^{*} =  0$ is also too small to be distinguished from \\
each other, which can be seen in the fourth and the \\
fifth column of Table 2, so the transition densities no \\
longer continue to be listed.}
\end{table*}

In the interior of a neutron star, $E_{F}(e)$ is commonly determined
by $\rho$, $Y_{e}$ and $B$.  In a given weakly quantizing strong
magnetic field ($B^{*} \leq 1$), $E_{F}(e)$ increases with matter
density $\rho$, though $Y_{e}$ decreases slightly with the nucleon
number $A$.  For a perfect crystal lattice with a single nuclear
species ($A, Z$), the magnetic field enhances the nuclear transition
density, causing a significant increase of $E_{F}(e)$. From Table 2,
we can infer that the experimental value of 0.95 MeV (see the first
row of Table 2) is evidently smaller than the theoretically correct
value (no less than 1 MeV) because when the electron kinetic energy
is comparable to its rest energy ($mc^2$), the electron is
relativistic.  Meanwhile, the experimental value of 2.61 MeV (see the
second row of Table 2) must be higher than the theoretical value (less
than 1.96 MeV), otherwise the calculated value 1.96 MeV is incorrect.

Solving Eq.(22) gives a useful special solution for $E_{F}(e)$
\begin{equation}
 E_{F}(e)= 34.9[\frac{Y_{e}}{0.05} \frac{\rho}{\rho_{0}}\frac{B}{B_{cr}}]
^{\frac{1}{4}}{\rm MeV}        (B^{*} \geq 1).
\end{equation}
From Eq.(23), we obtain the schematic diagrams of $E_{F}(e)$ vs. $B$
and $E_{F}(e)$ vs. $\rho$ as shown in Figure 1.

\begin{figure}[th]
\centering
 \vspace{0.5cm}
\subfigure[]{
    \label{evsb:a} 
    \includegraphics[width=7.2cm]{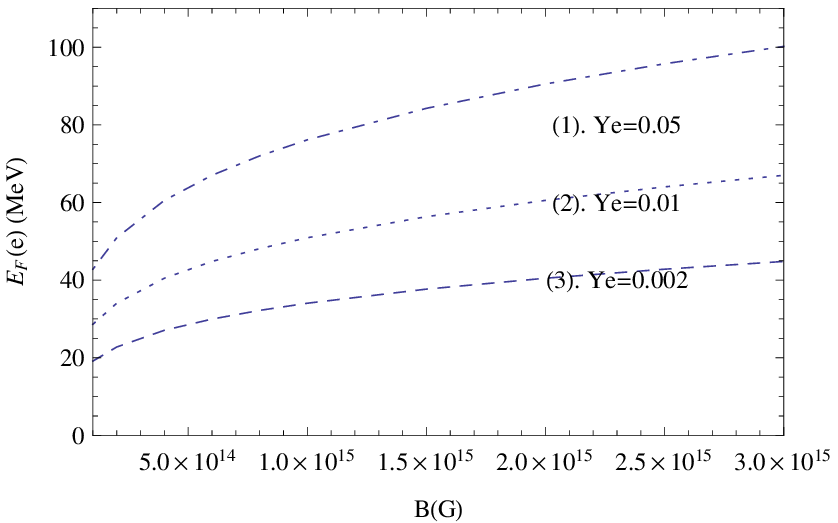}}
  \hspace{0.2mm}
  \subfigure[]{
    \label{evsb:b} 
    \includegraphics[width=7.0cm]{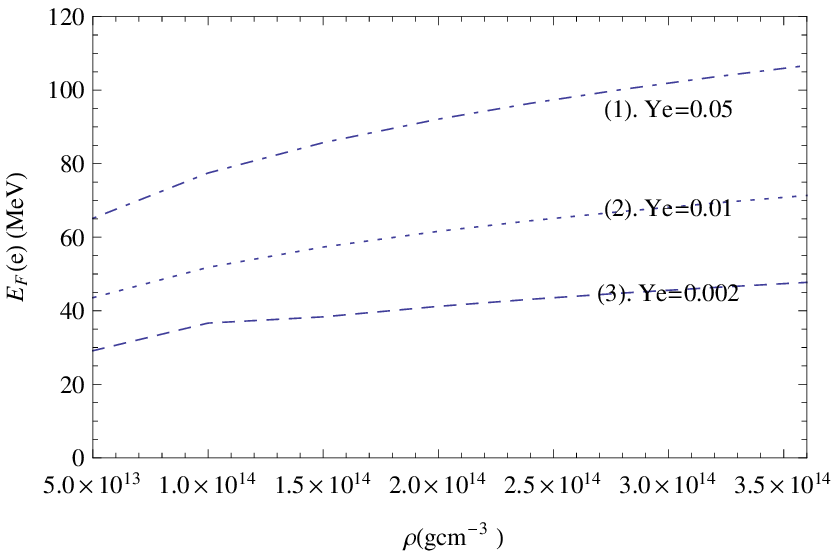}}
 \caption{Top, the electron Fermi energy dependence of the
 magnetic field strength when $\rho$ = 2.8 $\times 10^{14}$ g~cm$^{-3}$.
 The range of $B$ is (0.5$\times 10^{14}\sim$ 3.0$\times 10^{15}$)G.
 Bottom, the electron Fermi energy dependence of the matter density
 when $B$ = 3 $\times 10^{15}$ G.  The range of $\rho$ is (0.5 $\sim$ 3.6)
 $\times 10^{14}$ g~cm$^{-3}$.}
 \label{fig:Figure 1}
 \end{figure}

From Figure 1, it is clear that $E_{F}(e)$ increases with increasing
$B$ in the case of superhigh magnetic fields. We speculate that the
high Fermi energy of electrons could be from the release of the magnetic
energy according to our model.
\section{A dispute on the electron Fermi energy in intense magnetic fields}
This section is composed of three subsections.  For each
subsection we present different methods and considerations.
\subsection{Interpretations of high Fermi energy of electrons}

In this part, the possible interpretations of high $E_{F}(e)$ are given
as follows.

Firstly,  the electron Fermi energy increases with matter density.
In the interior of a magnetar, magnetic fields are closely
interconnected with matter density, and an extremely strong magnetic
field could exist in the depths of the star where matter density
$\rho \sim (10^{14}\sim 10^{15})$ g~ cm$^{-3}$ and $E_{F}(e)\geq$ 100
MeV \cite{yak01}, and the value of $Y_{e}$ is expected to be higher
than the mean value of $Y_{e}$ of a NS, which implies that $E_{F}(e)$
is also expected to increase.  For example, hyperon, pion condensates,
kaon condensates, quarks and nucleons with large $Y_{e}$ are expected
in the inner core of a NS \citep{tsu02,tsu09}, the maximum of the
inner core density  could exceed the transition density $\rho_{tr}$,
the value of $E_{F}(e)$ could be far larger than 100 MeV, accordingly.

Secondly, due to the existence of a weakly quantizing intense
magnetic field and a non-quantizing intense magnetic field in
the interior of a magnetar, $n_{max}$ could be large, which implies the
cyclotron energy of electrons is high, hence the electron Fermi energy
is also high.

Finally, in the case of field-free (or weak field), both $dp_{z}$
and $dp_{\perp}$ change continuously, the microscopic state
number in a volume element of phase space $d^{3}x d^{3}p$ is
$d^{3}x d^{3}p/h^{3}$.  In the presence of an intense magnetic
field, $dp_{z}$ changes continuously, whereas $dp_{\perp}$ is
not continuous and must obey the Landau relation: $(p_
{\perp}/ m_{e}c)^{2}= (2n + 1+ \sigma)B^{*}$, $n = 0, 1, 2, \cdots,
n_{max}(p_{z}, B^{*}, \sigma)$.  For a given $p_{z}$, there is a
maximum orbital quantum number $n_{max}(p_{z}, B^{*}, \sigma)\approx
n_{max}(p_{z}, B^{*})$.  In superhigh magnetic fields, an envelope of
these Landau cycles with maximum orbital quantum number $n_{max}
(p_{z}, B^{*})$($0\leq p_{z}\leq p_{F}$)will approximately form a
sphere, i.e. a Fermi sphere.  The number of states in the $x-y$
plane will be less than that when the magnetic field is absent.
For a given electron number density in a highly degenerate
state, the stronger the magnetic field, the larger the maximum
of $p_{z}$, hence the lower the number of states in the
$x-y$ plane according to the Pauli exclusion principle (each
microscopic state is only occupied by one electron ).  In
other words, when $B$ increases, both $n_{max}(p_{z},
B^{*})$ and the number of electrons in the $x-y$ plane decrease;
accordingly, the radius of the Fermi sphere $p_{F}$ expands,
which means that the electron Fermi energy $E_{F}(e)$ also
increases.  It should be noted that the higher the electron Fermi
energy, the more obvious the ` expansion ' of the Fermi sphere;
however, the majority of the momentum space in the Fermi sphere
is empty and unoccupied by electrons.

In a word, the stronger the magnetic field, the higher the electron
Fermi energy; the high Fermi energy of electrons could be supplied
by the release of the magnetic energy.
\subsection{A wrong conclusion on the electron Fermi energy}
The Fermi energy of the electrons is generally believed to decrease
with the increase of the magnetic field strength for ultrastrong
magnetic fields.  The reasons for this are as follows: The non-
relativistic Schr\"{o}dinger Equation for the electrons in a uniform
external magnetic field along the $z-$axis gives the electron energy
level
\begin{equation}
  E_{e}= p_{z}^{2}c^{2}/2m_{e}
 +(2n+1+\sigma)\hbar\omega_{B},
 \end{equation}
where $\hbar\omega_{B}= 2\mu_{e}B$, $\omega_{B}$ is the well-known
non-relativistic electron cyclotron frequency (c.f Page 460 of
Quantum Mechanics \citep{lan65}).  In the
direction perpendicular to the magnetic field, the energy of
electrons is quantized.  In the interval [$p_{z}$ $p_{z}+dp_{z}$]
along the magnetic field, for a non-relativistic electron gas, the
possible microstate numbers are given by
\begin{equation}
N_{pha}(p_{z})=\frac{eB}{4\pi \hbar^{2}}\frac{dp_{z}}{c}.
\end{equation}
Therefore, we obtain
\begin{equation}
N_{pha}=\int_{0}^{p_{F}}N_{pha}(p_{z})dz= \frac{eB}{4\pi
\hbar^{2}}\frac{E_{F}(e)}{c^2},
\end{equation}
where the solution of Eq.(24) is used (also c.f. Page 460 of
Quantum Mechanics \citep{lan65}).  In the light of the Pauli
exclusion principle, the electron number density should be equal
to its microstate density,
\begin{equation}
N_{pha}=n_{e}=\frac{eB}{4\pi \hbar^{2}}\frac{E_{F}(e)}{c^2}=
N_{A}\rho Y_{e}.
\end{equation}
From Eq.(27), it is easy to see $E_{F}(e) \propto B^{-1}$
when $n_{e}$ is given.  We then ask why such a phenomenon
exists.  After careful analysis, we find that the solution of
the non-relativistic electron cyclotron motion equation
$\hbar\omega_{B}$ is incorrectly(or unsuitably) applied to
calculate the energy state density in a relativistic degenerate
electron gas.  It's interesting to note that, in Page 12
of Canuto \& Chiu (1971), in order to evaluate the degeneracy
of the $n$-th Landau level $\omega_{n}$, they first introduce
the cylindrical coordinates ($p_{\perp},\phi$) where $\phi
=\arctan p_{x}/p_{y}$, and obtain an approximate relation
\begin{eqnarray}
&&\omega_{n}=(2\pi\hbar)\int_{0}^{2\pi}d\phi\int_{A<p_{\perp}^{2}<B}p_{\perp}dp_{\perp}\nonumber\\
&&=2\pi(2\pi\hbar)^{-2}\frac{(B-A)}{2}=\frac{1}{2\pi}(\frac{\hbar}{m_{e}c})^{-2}\frac{B}{B_{cr}},
\end{eqnarray}
where $A= m^{2}c^{2}\frac{B}{B_{cr}}2n$ and $B= m^{2}c^{2}
\frac{B}{B_{cr}}2(n+1)$\citep{can71}. The authors
stated clearly that this relation is valid only when $B$=0, in
other words, this relation is just an approximation in the case
of weak magnetic field ($B\ll B_{cr}$).  Surprisingly, this
relation has been misused for nearly 40 years since then. Even
in some textbooks on statistical physics, the statistical weight
is calculated unanimously by using the expression
\begin{equation}
 \frac{1}{h^2}\int dp_{x}dp_{y}= \frac{1}{h^2}\pi p_{\perp}^{2}
 \mid _{n}^{n+1}=\frac{4\pi m_{e}\mu_{e}B}{h^{2}}.
 \end{equation}
This expression will also cause the wrong deduction: $E_{F}(e)\propto
B^{-1}$, which is exactly the same as that from Eq.(27). This
wrong deduction is due to the assumption that the torus located between
the $n$-th Landau level and the $(n+1)$-th Landau level in momentum
space is ascribed to the $(n+1)$-th Landau level instead of using
Eq.(28).  Thus, the electron energy (or momentum) will change
continuously in the direction perpendicular to the magnetic field,
which is contradictory to the quantization of energy (or momentum) in
the presence of intense magnetic field.  Actually, the electrons are
relativistic and degenerate in the interior of a NS, so that Eq.(24)
and Eq.(26) are no longer applicable.  We therefore replace
Eq.(24) and Eq.(26) by Eq.(30) and Eq.(31), respectively.
\begin{equation}
  E_{e}^{2}= m_{e}^{2}c^{4}+p_{z}^{2}c^{2}
 +(2n+1+\sigma)2m_{e}c^{2}\mu_{e}B,
  \end{equation}
\begin{eqnarray}
&&N_{pha}=\frac{2\pi}{h^{3}}\int dp_{z}\sum_{n=0}^{n_{m}(p_z,\sigma,B^{*})}\sum g_{n}\nonumber\\
 &&\int \delta(\frac{p_{\perp}}{m_{e}c}-[(2n+1+\sigma)B^{*}]^{\frac{1}{2}})p_{\perp}dp_{\perp},
 \end{eqnarray}
where the Dirac $\delta$-function $\delta(\frac{p_{\perp}}{m_{e}c}
-[(2n+1+\sigma)B^{*} ]^{\frac{1}{2}})$ must be taken into account,
for otherwise we would  reach the wrong conclusion that $E_{F}(e)$
decreases with the increase of the field strength, $B$, in intense
magnetic fields ($B\gg B_{cr}$).
\subsection{Observations of magnetars}
It is widely supposed that the magnetic field is the main energy
source of all the persistent and bursting emission observed in
AXPs and SGRs \cite{dun92, mer08}.  Based on the observation up
to now (10 February 2011) of nine SGRs (seven confirmed) and
twelve AXPs (nine confirmed) at hand, a statistical investigation
of relevant parameters is possible.  All known magnetars are X-ray
pulsars with luminosities of $L_{X}\sim(10^{32}\sim 10^{36})$
erg~s$^{-1}$, usually much higher than the rate at which the star
loses its rotational energy through spin-down \cite{rea10}.  In
Table 3, the persistent parameters of sixteen confirmed magnetars
are listed in the light of observations performed in the last two
decades.
\begin{table}
\small
\caption{AXP/SGR persistent parameters. \label{tb3}}
\begin{tabular}{@{}crrr@{}}
\tableline
Name& \multicolumn{1}{c}{$ B$\tablenotemark{a}} &$L_{X}$&\multicolumn{1}{c}{$dE/dt$\tablenotemark{b}}\\
\tableline
SGR0526-66           & 5.6       & 1.4$\times10^{35}$       & 2.9$\times10^{33}$\\
SGR1806-20             & 24        &5.0$\times10^{36}$\tablenotemark{c}    & 6.7$\times10^{34}$ \\
SGR1900+14            &7.0       &(0.83$\sim$1.3)  &2.6$\times10^{34}$ \\
                      &          &  $\times 10^{35}$                               &         \\
SGR1627-41          & 2.2       & 2.5$\times 10^{33}$      & 4.3$\times10^{34}$\\
SGR0501+4516        & 1.9        & NO                       & 1.2$\times10^{33}$\\
SGR0418+5729          &$<0.075$   & NO                    &$<3.2\times10^{29}$\\
SGR1833+0832        &1.8        & NO                       & 4.0$\times10^{32}$\\
\tableline
CXOUJ0100          & 3.9        & 7.8$\times 10^{34}$      & 1.4$\times10^{33}$\\
1E2259+586         & 0.59        & 1.8 $\times 10^{35}$      &5.6$\times10^{31}$\\
4U0142+61          &1.3       & $>$5.3$\times10^{34}$    &1.2$\times10^{32}$\\
1E1841-045        &7.1       & 2.2$\times10^{35}$      &9.9$\times10^{32}$\\
1RXSJ1708          &4.7        & 1.9$\times10^{35}$      &5.7$\times10^{32}$\\
CXOJ1647\tablenotemark{t}     &1.6       & $2.6\times10^{34}$  &7.8$\times10^{31}$\\
1E\tablenotemark{t}1547.0-5408    &2.2     & 5.8$\times 10^{32}$      &1.0$\times10^{35}$\\
XTEJ\tablenotemark{t}1810-197   &2.1     &1.9$\times10^{32}$      &1.8$\times10^{33}$\\
1E\tablenotemark{d}1048.1-5937  &4.2      &5.4$\times10^{33}$      &3.9$\times10^{33}$\\
 \tableline
\end{tabular}
\tablenotetext{a}{
The sign `a' denotes that the surface dipolar magnetic \\
field of a pulsar can be estimated using its spin period,  \\
$P$, and spin-down rate, $\dot{P}$, by $B\simeq 3.2\times 10^{19}(P \dot{P})^{\frac{1}{2}}$ G, \\
where $P$ is in seconds and $\dot{P}$ is in secods/second;}
\tablenotetext{b}{
The sign `b' indicates:  A pulsar slow down with time \\
as its rotational energy is lost via magnetic dipolar ra- \\
diation, and the loss rate of a pulsar's rotational energy\\
is noted as $dE/dt$;}
\tablenotetext{c}{
The sign `c' denotes: from Thompson \& Duncan 1996;}
\tablenotetext{t}{The `t' symbol indicates: transient AXP;}
\tablenotetext{d}{The `d' symbol indicates: dim AXP.}
\tablecomments{All data are from the McGill AXP/SGR online \\
catalog of 10 Feb. 2011(http://www.physics. mcgill.ca/$^{\sim}$\\
pulsar/magnetar/main.html) except for $L_{X}$ of SGR1806 \\
-20. The units of $B$, $L_{X}$ and $dE/dt$ are $10^{14}$ G, erg~s$^{-1}$  \\
and erg~s$^{-1}$, respectively.}
\end{table}

From Table 3, we obtain the schematic diagram of magnetar's soft
X-ray luminosity as a function of magnetic field strength  as
shown in Figure 2(a). From Figure 2(a), it's obvious that magnetar's
soft X-ray luminosity increases with the increasing magnetic field
strength. Furthermore, we also obtain the schematic diagrams of
magnetar's soft X-ray luminosity as a function of the electron
Fermi energy  The stronger the magnetic fields, the higher the
electron Fermi energy becomes.  Thus, magnetar's soft X-ray
luminosity increases with the increasing the electron Fermi
energy as shown in Figure 2(b).

  \begin{figure}[th]
 \centering
   \vspace{0.5cm}
  \subfigure[]{
    \label{evsb:a} 
    \includegraphics[width=7.2cm]{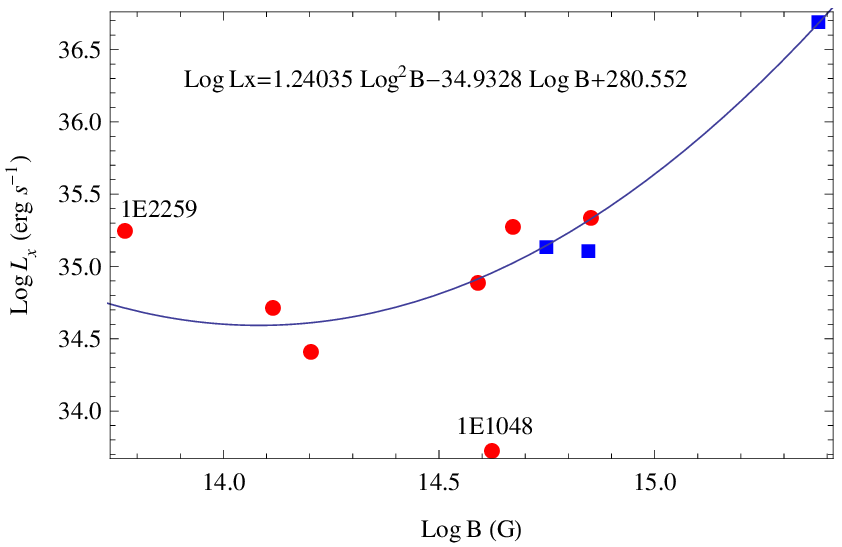}}
  \hspace{0.2mm}
  \subfigure[]{
    \label{evsb:b} 
    \includegraphics[width=7.0cm]{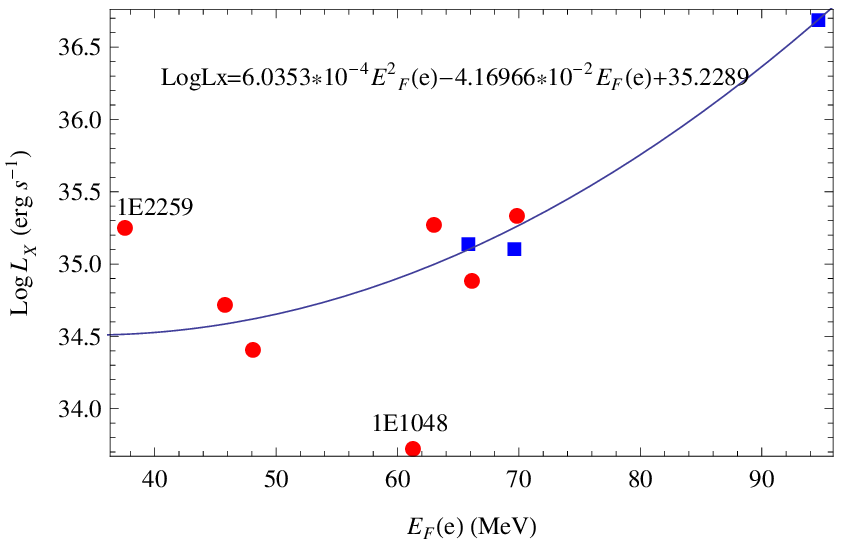}}
 \caption{Top, the fitted curve of $L_{X}$ vs. $B$ for magnetars. The range of
$B$ is (0.5$\times 10^{14}\sim$ 2.5$\times 10^{15}$)G.  Squares and circles
mark the values of variables corresponding to SGRs and AXPs, respectively.
Bottom, the fitted curve of $L_{X}$ vs. $E_{F}(e)$ for magnetars.  The range of
$E_{F}(e)$ is (36 $\sim$ 96) MeV, where $\rho$ =2.8$\times 10^{14}$
 g~cm$^{-3}$ and $Y_{e}$ = 0.05.}
 \label{fig:Figure 2}
 \end{figure}

It is particularly worth noting that  magnetars SGR 0501+4516, SGR
0418+5729 and SGR1833+0832 with no persistent soft X/$\gamma$-ray
fluxes observed need not be considered when fitting the curves in
Figure 2.  In addition, according to canonical magnetar model
\citep{dun92,dun96,tho96}, magnetar is a massive cooling isolated
neutron star with no accretion\citep{yak01}, and its persistent
soft X-ray luminosity shouldn't be less than its rotational energy
loss rate, $dE/dt$, so magnetars SGR 1627-41, 1E 11547.0-5408 and
XTEJ 1810-197 also should not be considered.  With respect to
AXPs 1E 2259+586 and 1E 1048.1-5937, their particular soft X-ray
luminosities are far from the fitted curved lines, shown in
Figure 2. The possible explanations are given as follows.

\begin{enumerate}
\item   The observed properties of 1E 2259+586 seem consistent
with the suggestion that it is an isolated pulsar undergoing a
combination of spherical and disk accretion \citep{whi84}. This
magnetar could be powered by accretion from the remnant of Thorne
-$\dot{Z}$ytow object(T$\dot{Z}$ \citep{van95}).

\item AXP 1E 1048.1-5937, discovered as a 6.4 s pulsar near
the Carina Nebula \cite{ste86}, is confirmed to be a dim isolated
pulsar with no mechanism to explain well its `abnormal' behaviors
including $L_{X}$.  Observations indicated a significant decline
in variation in its persistent soft X-ray luminosity.  For example,
between September 2004 and February 2011, $L_{X}$ decreased from
(1$\sim$ 2) $\times 10^{34}$ erg~s$^{-1}$ \citep{mer04} to 5.4
$\times 10^{33}$ erg~s$^{-1}$.
\end{enumerate}

However, in accordance with the traditional view on the electron
Fermi energy, the electron capture rate $\Gamma$ will also decrease
with increasing $B$ in ultrastrong magnetic fields.  If the electron
captures induced by field-decay are an important mechanism powering
magnetar's soft X-ray emission\citep{coo10}, then $L_{X}$ will also
decrease with increasing $B$, which is contrary to the observed data
in Table 3 and the fitting results of Figure 2.

\section{Conclusions}
In this paper, on the basis of the distribution of Landau levels
of electrons, we derive the formulae for electron Fermi energy
in ultrastrong magnetic fields.  We conclude that the stronger
the magnetic fields, the higher the electron  Fermi energy becomes.
However, the traditional viewpoint on electron Fermi energy will
be confronted with a severe challenge from our calculations of
$E_{F}(e)$.   If the magnetic field is the main energy source of
all the persistent and bursting emission observed in magnetars, this
article could be useful in studying the direct URCA processes,
neutrino emissions and neutron star cooling, etc.  It is expected
that that our assumptions and calculations can be used to compare with
observations in the future, to provide a deeper understanding of
the nature of the ultrastrong magnetic fields and soft X-ray in
magnetars.

\begin{acknowledgments}
We are very grateful to Prof. Qiu-He Peng, Prof. Zi-Gao Dai
and Prof. Yong-Feng Huang for their help in improving our presentation.
This work is supported by National Basic Research Program of China (973
Program 2009CB824800), Knowledge Innovation Program of The Chinese
Academy Sciences KJCX$_{2}$-YW-T09, West Light Foundation of CAS
(No.280802), Xinjiang Natural Science Foundation No.2009211B35, the
Key Directional Project of CAS and NSFC under projects 10173020,10673021,
10773005, 10778631,10903019 and 11003034.
\end{acknowledgments}

\end{document}